\documentclass{desyproc}
\newcommand{\be}{\begin{equation}}
\newcommand{\ee}{\end{equation}}

\begin{document}
\title{Differences between Axions and Generic Light Scalars in Laboratory Experiments}

\author{{\slshape Sonny Mantry$^1$\footnote{Speaker at the 10$^{th}$ PATRAS Workshop on Axions, WIMPs, and WISPs, 2014, CERN, Geneva, Switzerland.}, Mario Pitschmann$^{2}$ and Michael J. Ramsey-Musolf$^{3,4}$}\\[1ex]
$^1$Department of Physics, University of North Georgia, Dahlonega, GA, USA\\
$^2$Institute of Atomic and Subatomic Physics, Vienna University of Technology, Vienna\\ 
$^3$Amherst Center for Fundamental Interactions, Department of Physics, University of Massachussetts Amherst, Amherst, MA, USA\\
$^4$California Institute of Technology, Pasadena, CA, USA}

\contribID{familyname\_firstname}

\desyproc{DESY-PROC-2014-XX}
\acronym{Patras 2014} 

\maketitle

\begin{abstract}
It is well-known that electric dipole moment (EDM) constraints provide the most stringent bounds on axion-mediated macroscopic spin-dependent (SD) and time reversal and parity violating (TVPV) forces. These bounds are several orders of magnitude stronger than those arising from direct searches in fifth-force experiments and  combining astrophysical bounds on stellar energy loss with E\"otv\"os tests of the weak equivalence principle (WEP). This is a consequence of the specific properties of the axion, invoked to solve the Strong CP problem. However, the situation is quite different for generic light scalars that are unrelated to the strong CP problem. In this case, bounds from fifth-force experiments and astrophysical processes are far more stringent than the EDM bounds, for the mass range explored in  direct searches.
\end{abstract}

In this work~\cite{Mantry:2014zsa}, we consider the nature of constraints on macroscopic spin-dependent (SD) and T- and P-violating (TVPV) forces mediated by light scalar particles. In particular, we focus on differences between forces mediated by axions that solve the Strong CP problem and generic scalars that are unrelated to the Strong CP problem.  Here macroscopic forces are understood to have an interaction range  $r \gg 1$ \AA. For example, such a force can arise at the microscopic level through a coupling of a light scalar $\varphi$  with the light quarks $q=u,d$ 
\begin{eqnarray}
\label{axionquark}
{\cal L}_{\varphi qq} &=& \varphi \>\bar{q} \big(g_s^q+ig_p^q\gamma^5\big) q\ \ \ ,
\end{eqnarray}
which in turn can induce nucleon level couplings
\begin{eqnarray}
\label{axionquark}
\label{axionnucleon}
{\cal L}_{\varphi NN} &=& \varphi \>\bar{N}\big( g_s + ig_p\gamma^5 \big) N\ \ \ ,
\end{eqnarray}
where the nucleon-level couplings $g_{s,p}$ are related to the quark level couplings $g_{s,p}^q$ via nuclear matrix elements as determined by a matching calculation. For simplicity, we have assumed isoscalar couplings so that $g_{s,p}^u=g_{s,p}^d$ and ignored possible couplings to leptons. Such interactions give rise to a nucleon-nucleon monopole-dipole potential in the non-relativistic limit that has the form~\cite{Moody:1984ba}
\begin{eqnarray}
\label{potential}
V(r) &=& g_s g_p \>\frac{\vec{\sigma}_2\cdot \hat{r}}{8\pi M_2} \Big ( \frac{m_\varphi}{r} + \frac{1}{r^2}\Big )e^{-m_\varphi r}\ \ \ ,
\end{eqnarray}
where $\vec{\sigma}_2$ acts on the spin of the polarized nucleon and  $\hat r = \vec r/r$ is the unit vector from the unpolarized nucleon to the polarized nucleon. Direct searches in fifth-force experiments and astrophysical bounds on stellar energy loss, yield (or plan to yield) upper limits~ \cite{Adelberger:2006dh,Geraci:2008hb,Bordag:2001qi,Decca:2005qz,Nesvizhevsky:2007by,Abele:2003ga,Lamoreaux:1996wh,Klimchitskaya:2001zz,Horvat:2011jh,Bulatowicz:2013hf,Tullney:2013wqa,Chu:2012cf,Arvanitaki:2014dfa} on the product of couplings $g_s g_p$. A summary of various experiments  can be found in Ref.~\cite{Antoniadis:2011zza}. 

\begin{figure}[hb]
\centerline{\includegraphics[width=0.85\textwidth]{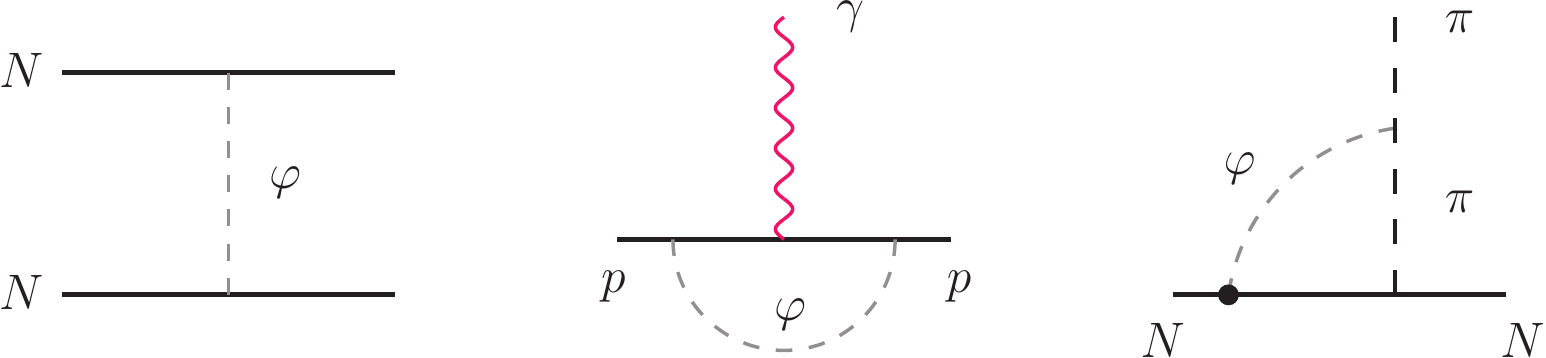}}
\caption{Example diagrams of $\varphi$-exchange that can contribute to nuclear EDMs. }
\label{Fig:MV}
\end{figure}

Since the nucleon-nucleon potential in Eq.~(\ref{potential}) is TVPV, it will induce non-zero electric dipole moments (EDMs) in nucleons and nuclei. The EDM for an elementary fermion arises from a term in the  Lagrangian of the form
$
{\cal L} = - i\> \frac{d}{2}\> \bar{\psi} \sigma^{\mu \nu} \gamma^5 \psi \>F_{\mu\nu}$,
which gives rise to the non-relativistic Hamiltonian of the form
$
H = -d \> \vec{E}\cdot \frac{\vec{S}}{S}$,
where $\vec{S}$ is the spin of the particle and $\vec{E}$ is the electric field.  For a non-zero value of $d$, TVPV or CP violation arises as a consequence of the CPT theorem and the time-reversal behavior of the interaction  $T(\vec{E}\cdot \vec{S})=-\vec{E}\cdot \vec{S}$. Current bounds from EDM experiments yield a bounds on the EDMs of the neutron  
$ |d_n| < 2.9 \times 10^{-13}$ e-fm~\cite{Baker:2006ts} and the diamagnetic mercury atom 
 $|d_{Hg}| < 2.6 \times 10^{-16}$ e-fm~\cite{Griffith:2009zz}. SInce the TVPV nucleon-nucleon potential in Eq.~(\ref{potential}) can induce non-zero EDMs, these EDM bounds translate into bounds on $g_sg_p$. Some examples of diagrams involving $\varphi$-exchange that contribute to nuclear EDMs and are proportional to $g_s g_p$ are shown in Fig.~\ref{Fig:MV}.\\ 
 
We are then led to ask the question of how bounds on $g_sg_p$ arising from fifth-force experiments compare with those arising from EDM constraints. It is well-known that when $\varphi$ is the axion($a$) \cite{Peccei:1977hh,Peccei:1977ur,Weinberg:1977ma,Wilczek:1977pj} that solves the strong CP problem, EDM constraints on $g_s g_p$ are several orders of magnitude more stringent~\cite{Rosenberg:2000wb} than those derived from fifth-force experiments. As we explain below, this is result of the unique properties of the axion that arise from the need to solve the Strong CP problem. However, we show that for the case when $\varphi$ is a generic scalar, unrelated to the Strong CP problem, the situation is quite different and the bounds from fifth-force experiments on $g_sg_p$ can be several orders of magnitude more stringent than those arising from EDM experiments. 

\section{Axion Scenario}
We can understand the differences in the bounds on $g_sg_p$ between axions and generic scalars due to the unique properties of the axion couplings and their connection to the Strong CP problem.  For the purposes of illustration, we consider QCD with one quark flavor. The terms relevant to the Strong CP problem are given by
\begin{eqnarray}
{\cal L} &=& \bar\theta\>\frac{\alpha_s}{16\pi}\> G^a_{\mu \nu}\tilde{G}^{a\mu\nu}  - m_q\>\bar qq\ \ \ .
\label{qcd-1}
\end{eqnarray}
The $\bar{\theta}$-term is the source of flavor-diagonal CP violation in QCD.  One can perform an axial $U(1)_A$ transformation so that the Lagrangian in Eq.~(\ref{qcd-1}) becomes
\begin{eqnarray}
\label{massCPV1}
{\cal L} &=& - m_q\cos\bar\theta\>\bar qq + m_q\sin\bar\theta\>\bar qi\gamma^5q\ \ \ ,
\end{eqnarray}
with strong CP violation now moved entirely into the quark mass terms. So far there has been no observable strong CP violation  and current EDM bounds require $|\bar\theta|< 10^{-10}$. This is the well-known Strong CP problem. The axion provides a solution~\cite{Peccei:1977hh,Peccei:1977ur,Weinberg:1977ma,Wilczek:1977pj} to the Strong CP problem by extending the Standard Model (SM) with new fields that are charged under a new anomalous $U(1)_{PQ}$ Peccei-Quinn symmetry, under which the SM fields are neutral. This symmetry can be used to completely rotate away the $\bar\theta$-term, thereby solving the Strong CP problem. However, in order to avoid additional light QCD degrees of freedom,  the Peccei-Quinn symmetry must be broken at a high energy scale $10^{9}\lesssim f_a \lesssim 10^{12}$ GeV. The axion field $\mathrm a(x)$ is the pseudo-Goldstone boson that arises from the spontaneously broken $U(1)_{PQ}$ symmetry. A $U(1)_{PQ}$ transformation results in the shifts
$
\bar{\theta} \to \bar{\theta} + 2\alpha\, ,  \frac{\mathrm  a(x)}{f_a} \to \frac{\mathrm a(x)}{f_a} -2\alpha ,
$
so that the combination  $\bar{\theta} + \frac{\mathrm a(x)}{f_a} $ is left invariant. Thus, in the low energy effective theory where all heavy degrees of freedom have been integrated and only the axion and the SM fields remain, all interactions involving the axion must be built out of this invariant combination. Thus, the low energy effective theory  axion interactions can be obtained by making the replacement
$
 \bar{\theta} \to \bar{\theta} + \frac{\mathrm  a(x)}{f_a}$
in the QCD Lagrangian. Making this replacement in Eq.~(\ref{massCPV1}) we get the effective Lagrangian
\begin{eqnarray}
\label{La2}
{\cal L}_a &=& - \cos\Big(\bar\theta +\frac{\mathrm  a}{f_a}\Big)\>m_q\>\bar qq + m_q\sin\Big(\bar\theta +\frac{\mathrm a}{f_a}\Big)\>\bar qi\gamma^5q\ \ \ , 
\end{eqnarray}
which makes manifest the couplings of the axion to the SM quark. The axion can acquire a non-zero expectation value due to strong interaction quark condensates so that we can write the axion field as
$
\mathrm a(x) = \langle \mathrm a\rangle + a(x)$ ,
where $a(x)$ denotes the axion field corresponding to excitations above the vev  $\langle \mathrm a\rangle$. This gives rise to a new induced $\bar\theta$-parameter $\theta_{\text{eff}}= \bar{\theta} + \frac{\langle \mathrm a\rangle}{f_a}$, so that the Lagrangian in Eq.~(\ref{La2}) now takes the form
\begin{eqnarray}
\label{La3}
{\cal L}_a &=& - \cos \Big (\theta_{\text{eff}} +\frac{a}{f_a} \Big )\>m_q\>\bar qq + m_q\sin\Big(\theta_{\text{eff}} +\frac{ a}{f_a}\Big)\>\bar qi\gamma^5q\ \ \ .
\end{eqnarray}
Non-perturbative QCD effects generate an axion potential via a non-zero quark condensate, given by
$
V\Big(\theta_{\text{eff}} + \frac{a}{f_a}\Big) =\>-\chi(0)  \cos \Big (\theta_{\text{eff}} +\frac{a}{f_a} \Big)$,
where $\chi(0)=-m_q \>\langle \bar qq \rangle$. Expanding the potential $V(\theta_{\text{eff}})$ around its minimum gives $V(\theta_{\text{eff}})\simeq \frac{1}{2}\chi(0) \theta_{\text{eff}}^2$. i.e. the minimum of the potential corresponds to $\theta_{\text{eff}}=0$, so that the dynamical relaxation of the ground state axion potential solves the strong CP problem.

The presence of higher dimensional CP-odd operators, like the quark chromo-electric dipole moment, can generate terms that are linear in $\theta_{\text{eff}}$ so that the minimum of the potential is shifted to a small but non-zero value of $\theta_{\text{eff}}$. This can occur via correlators of the type~\cite{Pospelov:2005pr}
$
\chi_{\text{CP}} (0) = -i\> \text{lim}_{k\to 0}\> \int d^4x \> e^{ik\cdot x} \langle 0 | T(G\tilde{G} (x), {\cal O}_{\text{CP}} (0)) | 0 \rangle$
so that the expanded potential now takes the form
$
V(\theta_{\text{eff}}) \simeq \chi_{\text{CP}}(0) \>\theta_{\text{eff}}  + \frac{\chi(0)}{2}\>\theta_{\text{eff}}^2$.
The potential  is now minimized at a non-zero value  $\theta_{\text{eff}}=-\chi_{\text{CP}}(0)/ \chi (0)$. Thus, the axion scenario can generate non-zero EDMs while still providing a dynamical mechanism to explain the small size of strong CP effects in QCD. Expanding the axion Lagrangian in Eq.~(\ref{La3}) in $\theta_{\text{eff}}$ and $a(x)$, we arrive at
\begin{eqnarray}
\label{La4}
{\cal L}_a &=&  \Big ( \frac{\theta_{\text{eff}} }{f_a} a-1 \Big )\>m_q\>\bar qq + \Big (\theta_{\text{eff}} + \frac{a}{f_a}\Big)\>m_q\> \bar qi\gamma^5q + \frac{m_q}{2f_a^2}\>a^2\>\bar qq 
 + \cdots\ \ \ .
\end{eqnarray}
The mass and the couplings of the axion to the quarks are now manifest
\begin{eqnarray}
\label{gasp}
g_{a,s}^q &=& \frac{\theta_{\text{eff}} m_q }{f_a}\,, \qquad g_{a,p}^q = \frac{m_q }{f_a}\,, \qquad m_a \simeq \frac{1}{f_a} |\chi (0)|^{1/2}\ \ \ ,
\end{eqnarray}
where $g_{a,s}^q$ and $g_{a,p}^q$ denote the scalar and pseudoscalar couplings respectively. Note that since the Peccei-Quinn symmetry breaking scale $f_a\gg |\chi (0)|^{1/2}$, the axion is very light and can mediate a macroscopic force. Note that based on the quark-level couplings in Eq.~(\ref{gasp}), the product of the nucleon-level couplings will be of the form
\begin{eqnarray}
\label{gs1gp2a}
g_s^q g_p^q \propto \theta_{\text{eff}}\> \frac{m_q^2}{f_a^2}\ \ \ ,
\end{eqnarray}
where the constant of proportionality will be determined by nucleon matrix elements. We can see that the size of the SD fifth-force is heavily suppressed by $m_q^2/f_a^2 \ll 1$.

However, note that the dominant contribution to the nucleon and nuclear EDMs arise from nuclear matrix elements of the CP-odd mass term $\theta_{\text{eff}}  m_q\> \bar qi\gamma^5q$ in Eq.~(\ref{La4}). This term does not suffer from the suppression factor $m_q^2/f_a^2 \ll 1$ that occurs in the context of fifth-force experiments via Eq.~(\ref{gs1gp2a}). In other words, the dominant effect that generates an EDM is independent of the product of couplings $g_sg_p$. Thus, the properties of the axion allow for a relatively large effect in EDMs and a heavily suppressed effect for fifth-force experiments.

Current EDM bounds require $\theta_{\text{eff}}\lesssim 10^{-10}$. Using this value along with $m_q\sim 1$ MeV and a Peccei-Quinn scale  $f_a \sim 10^9 - 10^{12} \text{ GeV}$, corresponding to the axion window, gives a bound on $g_sg_p$ for the axion as
 \begin{eqnarray}
 \label{eq:axionbound}
g_s g_p \propto \theta_{\text{eff}}\> \frac{m_q^2}{f_a^2} < 10^{-40} - 10^{-34}\ \ \ .
 \end{eqnarray}
For  a more detailed discussion we refer the reader to Ref.~\cite{Mantry:2014zsa}

\section{Generic Scalar Scenario}
The situation is quite different for a generic light scalar, unrelated to the strong CP problem. For a generic scalar, a non-zero nucleon or nuclear EDM is generated via the exchange of $\varphi$ through diagrams. Thus, unlike the case of axions, the value of a non-zero nuclear EDM is proportional to $g_sg_p$ and arises through diagrams of the type shown in Fig.~\ref{Fig:MV}. However, for a generic scalar the product of couplings $g_sg_p$ is unrelated to the Strong CP parameter $\theta_{\text{eff}}$ and are \textit{a priori} unrestricted free parameters. 

The computation of nuclear EDMs is a  highly non-trivial many-body problem  involving hadronic and nuclear effects (see Refs.~\cite{Engel:2013lsa,Pospelov:2005pr,Ginges:2003qt} for recent reviews). We do not attempt to carry out rigorous computations and instead only aim to provide order of magnitude estimates. In particular, we estimate the contribution to the mercury ($^{199}$Hg) EDM from a generic scalar $\varphi$. The dominant contribution will arise from the first diagram in Fig.~\ref{Fig:MV} that involves a tree-level exchange of $\varphi$, proportional to $g_sg_p$. However, we do not have the machinery to perform a many-body computation of this effect involving the spin-dependent potential in Eq.~(\ref{potential}). In order to provide an order of magnitude estimate, we use the result that the nuclear EDM $d_{Hg}$ is given in terms of the nuclear Schiff moment $S_{Hg}$ as~\cite{deJesus:2005nb,Griffith:2009zz} 
 \begin{eqnarray}
 \label{dHg}
d_{Hg}\simeq -2.8\times10^{-4}\>\frac{ S_{Hg}}{\text{ fm}^2}.\ \ \  
 \end{eqnarray}
The Schiff moment is a function of TVPV pion-nucleon couplings $S_{Hg}=g_{\pi NN}\>\Big(a_0\>\bar{g}_{\pi NN}^{(0)} + a_1\>\bar g_{\pi NN}^{(1)} +a_2\>\bar g_{\pi NN}^{(2)}\Big)\>e\text{ fm}^3$ in the Lagrangian
 \begin{eqnarray}
 {\cal L}_{\pi NN} &=& \bar{g}^{(0)}_{\pi NN}\> \bar{N}\tau^a N\pi^a + \bar{g}_{\pi NN}^{(1)}\> \bar{N}N\pi^0 + \bar{g}^{(2)}_{\pi NN}\> \big(\bar{N} \tau^a N \pi^a - 3 \bar{N} \tau^3 N \pi^0\big)\ \ \ ,
\end{eqnarray}
where $g_{\pi NN}\simeq 13.5$ and $ \bar{g}^{(0)}_{\pi NN}, \bar{g}^{(1)}_{\pi NN}, \bar{g}^{(2)}_{\pi NN}$ denote the isoscalar, isovector, and isotensor TVPV pion-nucleon couplings, respectively. We compute the third diagram in Fig.~\ref{Fig:MV} and interpret the result as a contribution to the TVPV pion-nucleon couplings. In particular, we find that only the isoscalar component receives a non-zero result
 \begin{eqnarray}
 \label{eq:gpiNNshift1} 
\delta \bar g_{\pi NN}^{(0)} \simeq \frac{1}{16\pi}\frac{m_\pi^2 + m_\pi m_\varphi + m_\varphi^2}{m_\pi + m_\varphi}\frac{g_Am_\pi^2}{90\>\rm{MeV}m_Nf_\pi}g_s g_p\ \ \ ,
 \end{eqnarray}
and is proportional to $g_sg_p$. We refer the reader to Ref.~\cite{Mantry:2014zsa} for further details on the computation. The resulting shift in the nuclear Schiff moment and the current experimental constraint  $|d_{Hg}| < 2.6 \times 10^{-16}$ e-fm~\cite{Griffith:2009zz}, translates into a bound on $|g_s g_p| \lesssim 10^{-9}$. As noted earlier however, the dominant shift to the nuclear EDM will arise from the first diagram in Fig.~\ref{Fig:MV}, corresponding to a tree-level $\varphi$-exchange between nucleons and we expect it to be about two orders of magnitude larger than the loop-suppressed diagram that generates the shift $\delta \bar g_{\pi NN}^{(0)}$. Thus, we expect an upper bound on $g_sg_p$ in the range
 \begin{eqnarray}
\label{eq:edmrange}
|g_s g_p| \lesssim \big[10^{-11},10^{-9}\big]\ \ \ .
 \end{eqnarray}
Thus, we see that the EDM bound on $g_sg_p$ is much weaker for a generic scalar, compared to the case of an axion as seen in Eq.(\ref{eq:axionbound}). 
\begin{table}
\renewcommand{\arraystretch}{1.5}
\begin{center}
\begin{tabular}{|c|c|c|}
\hline
 Properties & Axion ($a$)  &  Generic Scalar   ($\varphi$)  \\
\hline\hline
$\qquad${\rm   EDM Source}$\qquad$&$\qquad$TVPV quark mass  $\qquad$&$\qquad$ $\varphi$ -exchange $\qquad$ \\[-1ex]
&$\sim \bar{\theta} \>m_q\> \bar{q} \>i\gamma_5  \> q$ & \\[1ex]
\hline
$g_s $& $\sim \bar{\theta} \>\frac{m_q}{f_a}\propto \bar{\theta} \>m_a$  & arbitrary  \\[1ex]
\hline
$g_p $& $\sim \>\frac{m_q}{f_a}\propto \>m_a$  & arbitrary  \\[1ex]
\hline
$g_s g_p$& $\sim \bar{\theta} \>\frac{m_q^2}{f_a^2}\propto \bar{\theta} \>m_a^2$  & arbitrary \\[1ex]
\hline
\end{tabular}
\end{center}
\caption{Summary of differences between an axion ($a$) and a generic light scalar ($\varphi$) in terms of their couplings to quarks and contributions to non-zero EDMs.}
\label{axion-generic}
\end{table}
\section{Conclusion}
It is well-known that for axion-mediated macroscopic spin-dependent forces, the strongest bounds arise from electric dipole moment constraints. However, we have shown that for generic scalars, unrelated to the Strong CP problem, fifth-force experiments and astrophysical constraints provide bounds that are several orders of magnitude more stringent than those arising from electric dipole moment constraints. A summary of the main relevant  differences between axions and generic scalars is given in Table ~\ref{axion-generic}. Thus, these different experiments and observations can be complementary to each other in unraveling the true nature new macroscopic spin-dependent forces.
\section{Acknowledgments}
We acknowledge fruitful discussions with H. Abele, P. Chu, H. Gao, and T. G. Walker. 
This work was supported in part by: U. S. Department of Energy contracts  DE-AC02-06CH11357 (MP), DE-FG02-08ER41531 (MP and MJRM), and DE-SC0011095 (MJRM), the Wisconsin Alumni Research Foundation (MP and MJRM), University of North Georgia (SM), and the theoretical program on the contract I689-N16 by the Austrian ÒFonds zur F\"orderung der Wissenschaftlichen ForschungÓ (MP).


\begin{footnotesize}


\end{footnotesize}


\end{document}